\documentclass[paper,11pt]{JHEP}
\pdfoutput=1
\usepackage{graphicx}
\usepackage[centertags]{amsmath}
\usepackage{amsfonts} 
\usepackage{amssymb} \usepackage{amsthm}
\usepackage{bbold}
\allowdisplaybreaks[1]

\newcommand{\vev}[1]{\langle{#1}\rangle}
\def\one{{\rm 1\kern -.9mm l}}                             %

\def\beq{\begin{equation}}
\def\eeq{\end{equation}}
\def\beq{\begin{equation}}
\def\eeq{\end{equation}}
\def\beqa{\begin{eqnarray}}
\def\eeqa{\end{eqnarray}}
\newcommand{\eqa}{\begin{eqnarray}}
\newcommand{\ena}{\end{eqnarray}}
\newcommand{\eq}[1]{eq. (\ref{#1})}

\def\ii{\mathrm{i}}
\def\ee{\mathrm{e}}
\newcommand{\Z}{\mathbb{Z}}

\newcommand{\cD}{\mathcal{D}}

\newcommand{\cL}{\mathcal{L}}

\newcommand{\vp}{\! \stackrel{{\leftrightarrow}}{\partial}\!}
\newcommand{\dgb}{}
\newcommand{\dge}{}

\title{Line defects in the 3d  Ising model}
\author{
M. Bill\'o$^\diamondsuit$, M.  Caselle$^\diamondsuit$ , 
 D. Gaiotto$^\clubsuit$, F. Gliozzi$^\diamondsuit$ , M.Meineri$^\spadesuit$, 
R. Pellegrini$^\diamondsuit$\\
$^\diamondsuit$ Dipartimento di Fisica, Universit\`a di Torino\\
and Istituto Nazionale di Fisica Nucleare - sezione di Torino \\
Via P. Giuria 1, I-10125 Torino, Italy
\\
$^\clubsuit$ Perimeter Institute for Theoretical Physics,\\ 
31 Caroline St. N., 
Waterloo, Ontario, Canada N2L 2Y5
\\
$^\spadesuit$ Scuola Normale Superiore, Piazza dei Cavalieri 
7 I-56126 Pisa, Italy\\
and Istituto Nazionale di Fisica Nucleare - sezione di Pisa
}
\bigskip
\abstract{We investigate the properties of the twist line defect in the critical 3d Ising model using Monte Carlo simulations. In this model the twist line defect is the boundary of a surface of frustrated links or, in a dual description, the Wilson line of the $\Z_2$ gauge theory. We test the hypothesis that 
the twist line defect flows to a conformal line defect at criticality and 
evaluate numerically the low-lying spectrum of anomalous dimensions of the local operators which live on the defect as well as mixed correlation functions of 
local operators in the bulk and on the defect.}

\begin{document}

\section{Motivations and structure of the paper}
Conformal field theories are a crucial ingredient in both abstract and concrete investigations of quantum field theory. They control critical phenomena in condensed 
matter physics and the RG flows of generic quantum field theories 
\cite{{Cardy:1996xt},{Polyakov:1970xd}}. 
Through AdS/CFT, they even provide a framework to study quantum gravity 
\cite{Maldacena:1997re}. With some exceptions 
in two-dimensions, where the conformal group is infinite-dimensional, our understanding of generic conformal field theories is still quite poor. 
It is pretty clear that conformal symmetry is a rather restrictive constraint on a field theory: the very notion of universality in critical phenomena 
originates from the fact that there are relatively few ``simple'' conformal field theories which describe the infrared behaviour of a large variety of physical systems. 
The bootstrap program aims to use the constraint of conformal symmetry to classify and possibly even solve conformal field theories \cite{Ferrara:1973yt,Polyakov:1974gs}. Recent advances give 
some hope that the bootstrap strategy could be successful even in dimension higher than two, especially when combined with extra input from other numerical methods 
\cite{ElShowk:2012ht,ElShowk:2012hu}.

Given the importance of conformal symmetry, it is interesting to consider probes or modifications of a theory which preserve a large subgroup of the conformal group. 
A basic example would be a conformal boundary condition, i.e., a boundary condition which is left invariant by all conformal transformations which fix the position of the boundary, which form an $SO(D,1)$ subgroup of the $SO(D+1,1)$ conformal group of the bulk $D$-dimensional conformal field theory \cite{Cardy:1984bb,binderbook8}. 
More generally, we can consider the notion of a {\it conformal defect}: a $d$-dimensional defect in a $D$-dimensional conformal field theory which wraps a $d$-dimensional hyperplane (or a sphere) and is invariant under the $SO(d+1,1) \times SO(D-d)$ subgroup of the $SO(D+1,1)$ conformal group which preserves the 
hyperplane. Conformal defects should play an important r\^ole in studying the universal low-energy behaviour of any configuration where a quantum field theory is modified or excited 
in the neighbourhood of a large $d$-dimensional submanifold. It should also be possible to use conformal defects as theoretical tools to probe or constrain the properties of 
conformal field theories. This is definitely the case in two-dimensional models, or higher-dimensional superconformal field theory, and it may be the case in the context of the bootstrap program as well. See \cite{Liendo:2012hy} for a recent attempt in that direction. 

The purpose of this paper is to study numerically the properties of the twist line defect in the critical 3d Ising model, which coincides with the Wilson line defect in the 
dual $\Z_2$ lattice gauge theory formulation of the model \cite{Caselle:1996ii} . We will test the hypothesis that the twist line defect flows to a conformal line defect in the continuum limit and
determine the low-lying spectrum of anomalous dimensions of the operators which live on the defect. Our choice of theory and defect is dictated by theoretical and practical considerations. 
The 3d Ising model is currently the basic example of a 3d CFT amenable of a bootstrap analysis. The existence and properties of the twist line defects are intimately 
related to the $\Z_2$ flavor symmetry of the Ising model. Thus twist line defects are an example of a conformal defect whose existence 
may encode a crucial property of a CFT. On the practical side, Wilson line defects in the $\Z_2$ lattice gauge theory are well studied numerically in the confining phase as the endpoints 
for a confining string \cite{Caselle:2002rm}. This allows us to use well established numerical technology. 

The structure of the paper is the following. In section \ref{Section:ConformalDefects} we review the properties of conformal defects. In section \ref{Section:IsingMonodromy} we define the twist line defects in the Ising model, and the 
basic local operators we will use in correlation functions. In section \ref{Section:Results} we present our results and conclusions. 

\section{Conformal defects}
\label{Section:ConformalDefects}

In the study of quantum field theories, one typically encounters a variety of useful local probes and modifications of the underlying theory. 
The most common example are local operators, which probe or modify the theory at a point of space-time. Wilson and 't Hooft line operators in gauge theories
are classical examples of a probe or modification which extends along a line. Boundary conditions or domain walls modify the theory along 
a codimension one locus in space-time. General examples of $d$-dimensional defects in a $D$-dimensional field theory can be engineered, say, 
by adding to the Lagrangian of a $d$-dimensional field theory terms which depend on the degrees of freedom of the $D$-dimensional field theory \dgb restricted to the defect \dge. 

If we consider a $D$-dimensional field theory invariant under the $ISO(D)$ Poincar\'e group and flow to the far infrared, 
we typically expect to end up with a conformal field theory, possibly trivial, topological or free, i.e. a theory which is invariant under 
the $SO(D+1,1)$ group of conformal transformations. In a similar fashion, we can consider a modification or a massive excitation of the field theory localized near a 
$d$-dimensional hyperplane and preserving the $ISO(d) \times SO(D-d)$ subgroup of the Poincare group which fixes the hyperplane. 
As we flow to the infrared, we do not expect the modification to affect the critical properties of the theory in the bulk. We can thus hope that the 
localized modification will either disappear or flow to a ``conformal defect'', i.e. a defect of the conformal field theory, preserving the $SO(d+1,1) \times SO(D-d)$
subgroup of the conformal group which fixes the hyperplane. 

A priori, the result of the RG flow is expected to have scale invariance, but full conformal invariance is a stronger constraint. For unitary quantum field theories, 
scale invariance is expected on general grounds to imply conformal invariance \cite{Polchinski:1987dy}. It is not known if a similar result holds for scale invariant defects in a CFT as well. 
See \cite{Nakayama:2012ed} for some recent work on the subject. Indeed, one basic motivation for the present work was to check conformal invariance for a simple example of defect in a 
non-supersymmetric 3d CFT.  

Although conformal defects are somewhat analogous to local operators, there are some important differences. 
The most obvious difference is that while the set of local operators of a field theory is naturally 
part of the definition of what the theory is, the set of higher-dimensional defects which can be inserted in a given conformal field theory can be enormous, 
morally as large as the set of $d$-dimensional conformal field theories. For example, a generic superconformal boundary condition in $N=4$ 4d SYM theory
can be engineered at weak coupling by gauging a flavor symmetry of a generic three-dimensional $N=4$ SCFT. 
One may think that the possibility of considering such a variety of conformal defects is somewhat artificial, and that 
simple modifications of the theory in the UV will lead to a small class of simple defects in the IR. This expectation is incorrect:
a very simple defect in the UV theory may acquire a very intricate IR dynamics,
due to ``edge excitations'' of the bulk theory. Very little is known about possible constraints on how RG flow in the bulk may affect the 
degrees of freedom living at a defect. 

If the bulk theory is a strongly-coupled CFT in the IR, there is really no well-defined separation 
between degrees of freedom at the defect, and bulk degrees of freedom. 
The closest analogue to studying a ``defect conformal field theory'' is to look at the set of 
local operators which live at the defect. These local operators have many properties in common with operators in a $d$-dimensional CFT
with $SO(D-d)$ flavor symmetry, with one important exception: such a $d$-dimensional CFT would have a protected stress-tensor operator
of conformal dimension $d$. In general, there is no such ``defect stress tensor'' available as a local operator at the defect. 

On the other hand, every conformal defect should support a ``displacement operator'', which we will denote as $D^i$, $i = 1, \cdots, D-d$, 
which has dimension $d+1$ and transforms as a vector under the $SO(D-d)$ group of rotations around the defect. 
Intuitively, the displacement operator 
is something which can be added to the Lagrangian of the theory in order to displace the defect in the normal direction, 
much as the stress tensor is something which can be added to the Lagrangian in order to deform the metric of space-time. As we can deform the shape of a defect by a local diffeomorphism, 
there should be a relation between the displacement operator and the bulk stress-tensor. 

This relation can be made precise: the displacement operator controls the breaking of translation symmetry normal to the defect, and thus enters the stress-tensor Ward identities:
\begin{equation}
\partial_\mu T^{\mu i} = D^i \prod_j \delta(x^j)~,
\end{equation}
where we denote the $D$-dimensional indices with Greek letters such as $\mu$ and the
$D-d$ transverse indices 
 with latin letters such as $i,j,k$.
This Ward identity makes the protected quantum numbers of $D^i$ manifest. 
Notice that the Ward identity fixes the normalization of $D^i$, and thus the numerical coefficient in the two-point function 
\begin{equation}
\langle D^i(x) D^j(0) \rangle = \frac{C_D\, \delta_{ij}}{|x|^{2d+2}}
\end{equation}
is an intrinsic property of the defect. Intuitively, ``simple'' defects will have a small two-point function coefficient. For example, a trivial or topological defect has $C_D=0$.

From the perspective of conformal bootstrap, the correlation functions of the bulk CFT can be computed from the knowledge of the spectrum of bulk local operators and of the coefficients of three-point functions, whose functional form is determined by conformal symmetry. In a similar fashion, correlation functions of defect local operators can be computed from the knowledge of the spectrum of defect local operators and the coefficients of three-point functions of defect local operators. 
On the other hand, mixed correlation functions of local operators in the bulk and on the defect require one extra piece of information: the bulk-to-defect pairing, i.e. the coefficient of two-point functions involving one bulk operator and one boundary operator. The functional form of such two-point functions is also 
fixed by conformal invariance: we can use a conformal transformation to send the defect operator at infinity, and then use scale transformations, translations and rotations to move the bulk operator at whatever location in space-time we want to use as a reference point. 
For example, the correlation function of a scalar bulk local operator ${\cal O}$ and a scalar defect local operator $\mathfrak{o}$ takes the form 
\begin{equation}
\langle {\cal O}(x) \mathfrak{o}(0) \rangle = C^{\cal O}_\mathfrak{o} |x^\mu|^{-2 \Delta_\mathfrak{o}} |x^i|^{\Delta_\mathfrak{o} -\Delta_{\cal O}}
\label{mixed}
\end{equation}
where $|x^i|$ is the distance from the defect, and $|x^\mu|$ the distance from the origin. Indeed, when the defect local operator is sent to infinity, 
the correlation function depends only on the transverse distance, as $|x^i|^{\Delta_\mathfrak{o} -\Delta_{\cal O}}$, and then an inversion centered on the origin 
gives the general correlator. 

An alternative, useful point of view is to consider the possible OPE expansions available in the system: two bulk operators close to each other can be expanded 
as a sum of bulk local operators sitting at an intermediate location, two defect local operators close to each other can be expanded 
as a sum of defect local operators sitting at an intermediate location on the defect, and a single bulk local operator near the defect can be expanded into a 
sum of defect local operators. The latter bulk-to-defect OPE is a good way to use knowledge about the bulk local operators to learn about the possible defect local operators. The OPE expansion of a bulk operator cannot be empty, otherwise the correlation functions involving the defect and that operators would be all zero. For example, if the defect preserves some flavor symmetry, there must be a defect local operator for each representation of the flavor symmetry for which a bulk local operator exists. 
\dgb 
\dge

In this paper we study a conformal defect which belongs to a special class of monodromy defects. Monodromy defects can be defined in a CFT which is equipped with a 
flavor symmetry group $G$. We will focus on the case of discrete $G$, but most of our considerations apply to a continuous flavor group as well. 
If a conformal field theory has a flavor symmetry, we can define a trivial class of topological domain walls ${\cal D}_g$ associated to elements of the flavor group $g$ as follows: a correlation function in the presence of the domain wall 
is equal to the same correlation function without the domain wall, but with all local operators 
on one side of the wall transformed according to $g$. 

A monodromy defect is defined as any codimension $2$ conformal defect on which a ${\cal D}_g$ 
domain wall can end. Local operators which transform non-trivially under $g$ will be multi-valued around the monodromy defect. In particular,
this means that the OPE of a bulk operator charged under $g$ will contain defect local operators with fractional spin under the $SO(2)$ group of transverse rotations. 
For example, if $\phi$ is the angular coordinate around the defect, $r$ the radial coordinate in the plane perpendicular to the defect, and $G = \Z_2$, 
a $\Z_2$ odd operator ${\cal O}$ of conformal dimension $\Delta$ will have OPE
\begin{equation}
{\cal O}(r, \phi) \sim \sum_{n, a} \ee^{\ii (n+\frac{1}{2}) \phi} r^{\Delta_a - \Delta} \mathfrak{o}_{n+\frac{1}{2},a}
\end{equation}
involving defect local operators $\mathfrak{o}_{n,a}$ of conformal dimension $\Delta_a$ and half-integral $SO(2)$ spin $s= n+1/2$.

\section{A monodromy defect in the Ising model}
\label{Section:IsingMonodromy}
Consider the Ising model on a cubic lattice. The Ising model has a $\Z_2$ flavor symmetry which flips the spin at each site. There is an obvious realization of a $\Z_2$ topological domain wall in the theory: consider some hypersurface $S$, and flip the sign of the spin-spin interaction for edges which cross $S$. 
If $S$ is closed, or extends to infinity, we can simply flip all the spins on one side of $S$, and recover the standard Hamiltonian for the Ising model. 
Similarly, we can deform $S$ to a different hypersurface $S'$, by flipping all the spins in the region  between $S$ and $S'$.
On the other hand, if $S$ has a boundary,  which will generally consist of a codimension $2$ locus $L$ which does not cross any edges of the lattice, 
the location of $L$ is meaningful, and we obtain a lattice realization of a monodromy defect. Any choice of $S$ which is bounded by the same locus $L$
defines the same monodromy defect. 

In the two-dimensional Ising model, the monodromy defect is a local operator, which goes in the continuum limit to the disorder operator $\mu$. 
The OPE between the disorder operator $\mu$ and the spin operator $\sigma$ gives a spin $1/2$ operator: the free fermion $\psi$ hidden in the 2d Ising model.
Notice that in the context of the 2d Ising model, the free fermion still sits at the end of the topological $\Z_2$ domain wall, which in the language of RCFT 
is the topological domain wall labelled by the energy operator $\epsilon$. 

In the three-dimensional Ising model, which is the focus of this paper, the monodromy defect is a line operator. The 3d Ising model has a dual description as a $\Z_2$ gauge theory, 
in which the monodromy defect is a very fundamental object, i.e. the Wilson loop, and has no topological domain wall attached to it. 
In the $\Z_2$ gauge theory description, on the other hand, the spin operator is essentially a monopole operator, and sits at the end of a topological line defect. The topological line defect acquires a minus sign when crossing the Wilson line operator. This implements the anti-periodicity of the spin operator around the Wilson loop operator. 
This is analogue to the behavior of a fundamental Wilson loop and a monopole of minimal charge in a 3d gauge theory based on the $su(2)$ Lie algebra. 
The fundamental Wilson loop is allowed if we pick an $SU(2)$ gauge group, the basic monopole is allowed if we pick an $SO(3)$ gauge group.
If we try to include both operators in correlation functions, the monopole operator will be anti-periodic around the Wilson loop, and we will to place either of the two at the end of some topological defect which keeps track of the antiperiodicity. 

After inserting the monodromy line defect in the lattice Ising model, we can tune the interaction strength to make the bulk theory critical, and flow to the far infrared. 
As the spin operator is anti-periodic around the defect, the line defect can hardly disappear in the IR. It is natural to conjecture that it will flow to a conformal line defect. The operators on the monodromy line defect will be labelled by their integral or half-integral spin $s$ and conformal dimension $\Delta$. The displacement operator gives rise to local operators $D = D^1 + i D^2$ of spin $1$, and $\bar D = D^1 - i D^2$ of spin $(-1)$, both of dimension $2$. 

It is easy to argue that in a theory of a free scalar field, the OPE of the scalar field with a monodromy defect would contain spin $s=n+1/2$ primary operators of dimension $|s|+1/2$. Indeed, we can consider a general OPE 
\begin{equation}
\phi(x) \sim \sum_a f_a(z, \bar z) \mathfrak{o}_a + \cdots
\end{equation}
where $z=x^1 + i x^2= r\, \ee^{\ii\,\phi}$ is a complex coordinate in the plane orthogonal to the defect (we are taking the defect line along the $x^3$ direction), $\mathfrak{o}_a$ the primaries on the defect, and the ellipsis indicates descendants (derivatives) of the primaries. 
If we apply the free field equation of motion on the OPE and look at the coefficients of primaries, we can ignore derivatives along the defect, 
which give descendants. Thus the OPE coefficients must be harmonic functions of $z, \bar z$, and the OPE must take the form 
\begin{equation}
\phi(x) \sim \sum_n \left( \bar z^{n+1/2} \mathfrak{o}_{n+1/2} + c.c \right) +  \cdots
\end{equation}

The Ising model is quite close to the theory of a single free scalar field, at least as far as conformal dimensions are concerned. Thus 
we expect the OPE of the spin operator with the defect to be dominated by a spin $1/2$ operator $\psi$ of dimension close to $1$, 
\begin{equation}
\sigma(x) \sim r^{\Delta_\psi - \Delta_\sigma} \left[ \ee^{-\ii \phi/2} \psi + \ee^{\ii \phi/2} \bar \psi \right] + \cdots
\end{equation}
and that the leading contribution in a spin $s=n+1/2>0$ sector will be an operator of dimension close to $n+1$. 

On the other hand, the OPE of the energy operator $\epsilon$ with the defect involves operators of even spin, and thus should 
be dominated by the identity operator, and possibly the displacement operator, which we expect to be the operator of lowest dimension in the $s=1$ sector:
\begin{equation}
\epsilon(x) \sim r^{- \Delta_\epsilon} \mathrm{1} + r^{2- \Delta_\epsilon} 
\left[ \ee^{-\ii \phi} D + \ee^{\ii \phi} \bar D \right] + \cdots
\end{equation}
In a free scalar theory, defect local operators of integral spin could be built as bilinears of the scalar field modes, $\mathfrak{o}_{n+1/2}\mathfrak{o}_{m+1/2}$,
of dimension equal to $|s|+1$. We thus expect the Ising model to also include defect local operators of integral spin $s$ and conformal dimension close to $|s|+1$.
It may be possible to understand these ``Regge trajectories'' of defect local operators in terms of the approximate higher spin symmetry expected to hold in the 3d Ising model.

\subsection{The set-up}
Let us now describe in some more detail the realization of monodromy defects in the 3d critical Ising model on a cubic lattice with periodic boundary conditions.

\begin{figure}
\begin{center}

\begingroup
  \makeatletter
  \providecommand\color[2][]{%
    \errmessage{(Inkscape) Color is used for the text in Inkscape, but the package 'color.sty' is not loaded}
    \renewcommand\color[2][]{}%
  }
  \providecommand\transparent[1]{%
    \errmessage{(Inkscape) Transparency is used (non-zero) for the text in Inkscape, but the package 'transparent.sty' is not loaded}
    \renewcommand\transparent[1]{}%
  }
  \providecommand\rotatebox[2]{#2}
  \ifx\svgwidth\undefined
    \setlength{\unitlength}{300pt}
  \else
    \setlength{\unitlength}{\svgwidth}
  \fi
  \global\let\svgwidth\undefined
  \makeatother
  \begin{picture}(1,0.53539783)%
    \put(0,0){\includegraphics[width=\unitlength]{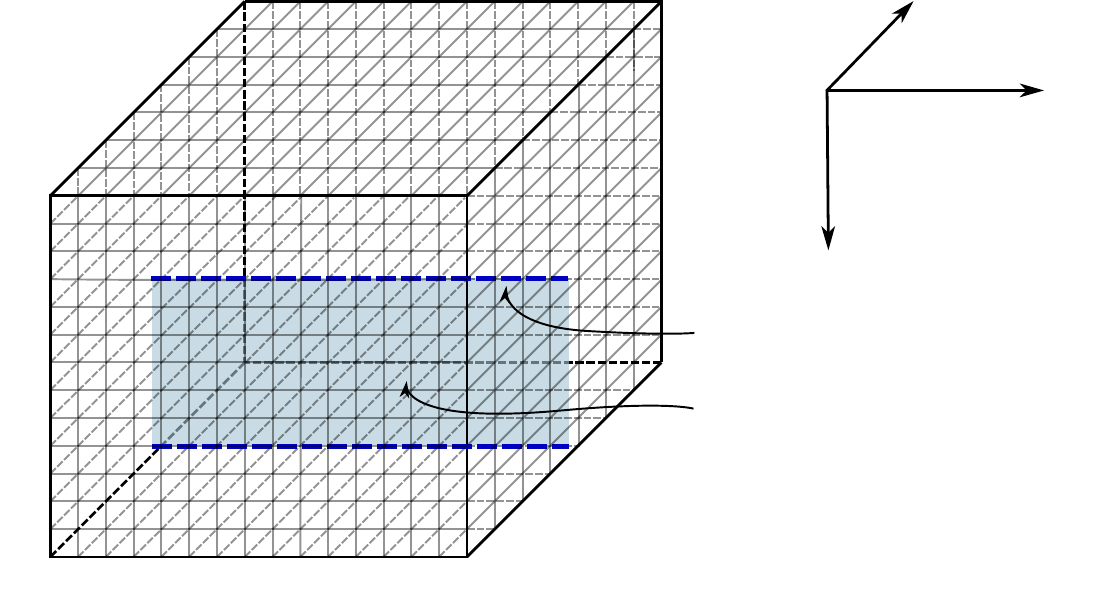}}%
    \put(0.90288564,0.41729222){\makebox(0,0)[lb]{\smash{$x^3$}}}%
    \put(0.81613863,0.5089434){\makebox(0,0)[lb]{\smash{$x^2$}}}%
    \put(0.75498003,0.33238632){\makebox(0,0)[lb]{\smash{$x^1$}}}%
    \put(0.23264388,0.00438318){\makebox(0,0)[lb]{\smash{$\ell$}}}%
    \put(-0.00247555,0.19848236){\makebox(0,0)[lb]{\smash{$\ell$}}}%
    \put(0.0879198,0.45812003){\makebox(0,0)[lb]{\smash{$\ell$}}}%
    \put(0.62614174,0.22766219){\makebox(0,0)[lb]{\smash{Monodromy line $L$ }}}%
    \put(0.62665945,0.16083326){\makebox(0,0)[lb]{\smash{Domain wall $S$}}}%
  \end{picture}%
\endgroup
\end{center}
\caption{In our set-up, the domain wall $S$ is the surface across which the links are frustrated. It ends on two defect lines. We will mostly consider the proximity of one of such lines, which 
we take to be aligned with the $x^3$ axis.}
\label{fig:setup}
\end{figure}
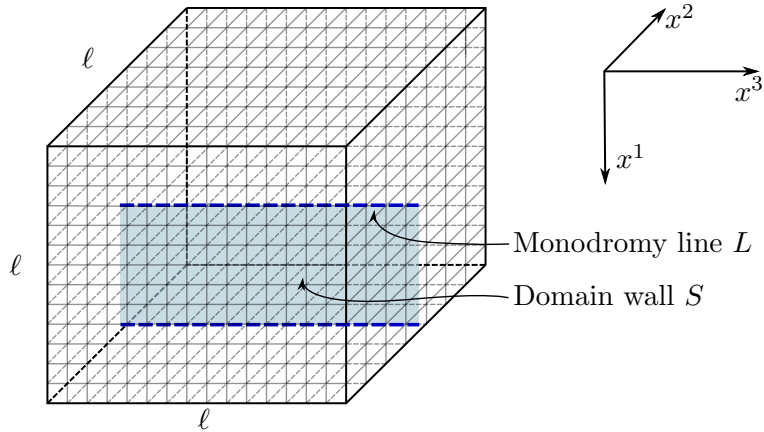

The partition function of the model on a cube of side $\ell$ is
\begin{equation}
 Z=\sum_{\{\sigma_{x}\}} \exp\left[-H\left(\{\sigma_{x}\}\right)\right]~,
\end{equation}
where the sum runs over all the spin configurations. The Hamiltonian reads
\begin{equation}
H\left(\{\sigma_{x}\}\right)=-\beta_{c} \sum_{\langle x y \rangle} J_{\langle x y \rangle}\, 
\sigma_{x} \sigma_{y}~,
\end{equation}
where the sum runs over nearest-neighbor sites and the $\Z_2$ 
variables $\sigma_{x}$ are defined on the sites. 

The coupling $ \beta_{c} $ is set to the best known critical value  $0.22165455$ \cite{Deng:2003wv}.
In the following we shall compare our results with the existing
estimates for the spin and energy critical dimensions. The most precise estimates for these quantities are  
$\Delta_\sigma=0.51813(5)$ and $\Delta_\epsilon=1.41275(25)$ from Monte Carlo simulations \cite{Martin2010} and 
$\Delta_\sigma=0.51819(7)$ and $\Delta_\epsilon=1.4130(4)$ from Strong Coupling Expansions \cite{phi4pisa}. A conservative combination of these results gives the two
values $\Delta_\sigma=0.5182(2)$ and $\Delta_\epsilon=1.4130(5)$
 which we shall use in the following.

As anticipated in Section \ref{Section:IsingMonodromy}, the position of the monodromy defects is encoded in the sign of the couplings of the spin-spin interaction $J_{\vev {xy}}$. We set these couplings
everywhere to $ +1 $, except on 
the bonds that intersect a surface $S$ joining two defect lines on 
the dual lattice  as depicted in Fig. \ref{fig:setup}, for which we choose $J_{\vev {x y} }=-1$.

On a finite lattice it is not possible to define a single 
straight defect; the simplest choice is to put a couple of defects as 
far as possible from each other and measure the observables in the 
neighborhood of one of them; if the size of the lattice is large enough the 
distant  defect lines do not disturb the measure. 
In some cases, in the correlation functions involving local operators on the bulk and on the defect the presence of other defect lines, including the copies generated by the periodic boundary conditions, cannot be neglected.%

Our aim is to realize on the lattice the lowest dimensional operators $\mathfrak{o}$ living on the monodromy defect (this will be the focus of the next section) and to compute their anomalous dimensions by means of Monte Carlo simulations. We will also study some correlators of local operators in the bulk and on the 
defect, as discussed in Eq. (\ref{mixed})
 
As a basic update algorithm we chose the standard Metropolis algorithm with multi-spin coding technique. Our version of this method is able to update
64 independent lattices in parallel on a simple desktop machine. 
It is important to notice that the defect breaks the translational symmetry 
in the two transverse directions and it would be a waste of CPU time to update the 
whole lattice before every measure. A simple way to speed up the simulation 
is to update sub-lattices of decreasing transverse dimensions centered around 
the defect in a hierarchical way \cite{Caselle:2002ah}. In order to avoid finite size effects 
the lattice size $ \ell $, in all the directions, has to be large enough;
it turns out that $ \ell=70 $ is adequate for computations involving even spin  operators,
 and $ \ell=120 $ is adequate for odd ones.

\subsection{Representations of the lattice symmetry group}
\label{representations}

As discussed in section \ref{Section:IsingMonodromy}, in a continuum theory with a monodromy line, defect local operators have definite scale dimension and $\mathrm{SO}(2)$ spin. We will realize some of 
such operators on the lattice in terms of spin operators sitting close to the monodromy line. 
These realizations are classified, rather than by their $\mathrm{SO}(2)$ spin, 
by the representation of the corresponding discrete symmetry on the lattice.  

\begin{figure}
\begin{center}

\begingroup
  \makeatletter
  \providecommand\color[2][]{%
    \errmessage{(Inkscape) Color is used for the text in Inkscape, but the package 'color.sty' is not loaded}
    \renewcommand\color[2][]{}%
  }
  \providecommand\transparent[1]{%
    \errmessage{(Inkscape) Transparency is used (non-zero) for the text in Inkscape, but the package 'transparent.sty' is not loaded}
    \renewcommand\transparent[1]{}%
  }
  \providecommand\rotatebox[2]{#2}
  \ifx\svgwidth\undefined
    \setlength{\unitlength}{250pt}
  \else
    \setlength{\unitlength}{\svgwidth}
  \fi
  \global\let\svgwidth\undefined
  \makeatother
  \begin{picture}(1,0.66082794)%
    \put(0,0){\includegraphics[width=\unitlength]{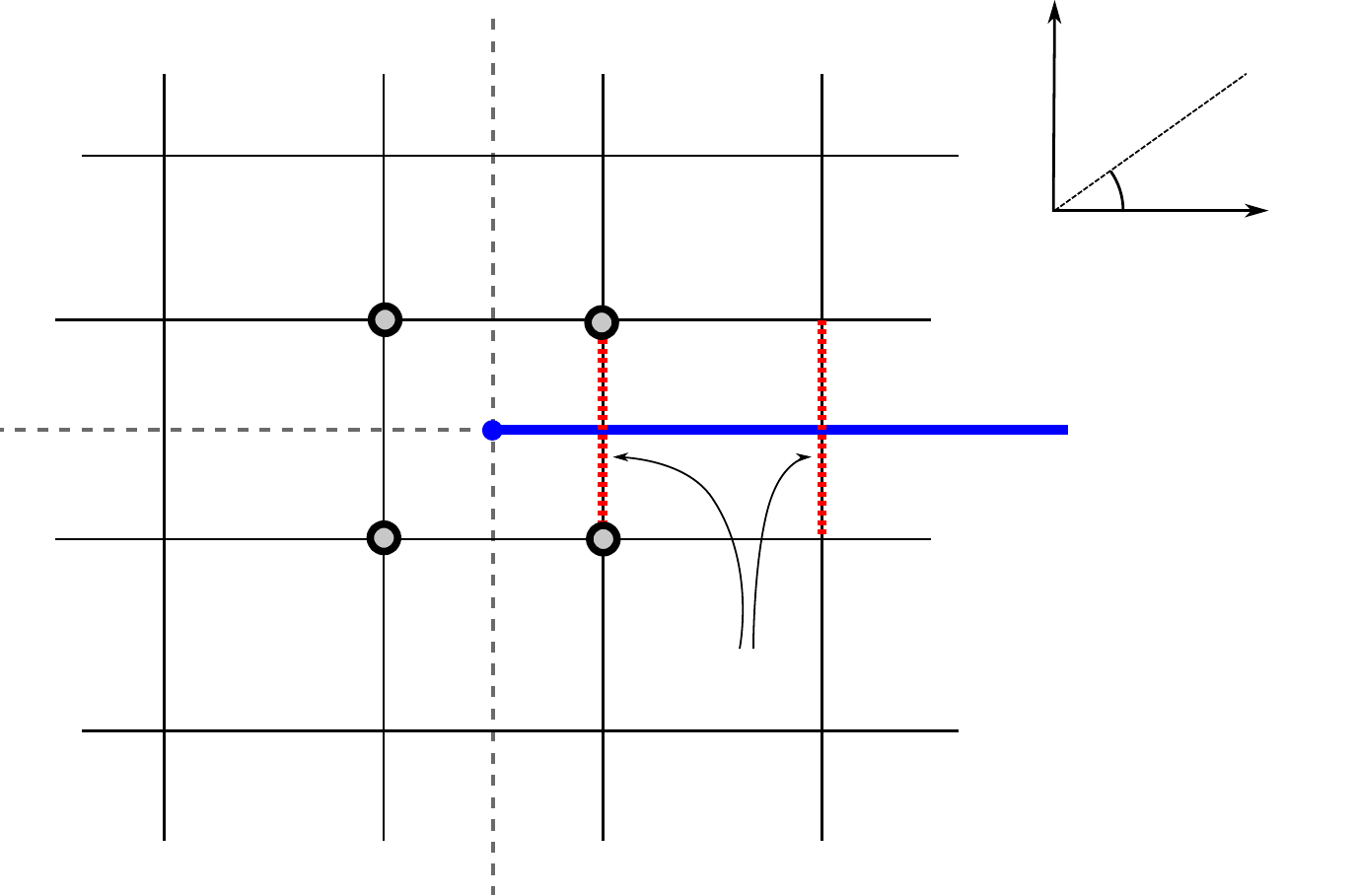}}%
    \put(0.45152578,0.22688754){\makebox(0,0)[lb]{\smash{$\sigma_1$}}}%
    \put(0.44950534,0.4482392){\makebox(0,0)[lb]{\smash{$\sigma_4$}}}%
    \put(0.23422493,0.45025962){\makebox(0,0)[lb]{\smash{$\sigma_3$}}}%
    \put(0.23118933,0.22688754){\makebox(0,0)[lb]{\smash{$\sigma_2$}}}%
    \put(0.50308664,0.15513237){\makebox(0,0)[lb]{\smash{frustrated links}}}%
    \put(0.6496026,0.36030565){\makebox(0,0)[lb]{\smash{$S$}}}%
    \put(0.31708745,0.35929043){\makebox(0,0)[lb]{\smash{$L$}}}%
    \put(0.71357048,0.63978341){\makebox(0,0)[lb]{\smash{$x^2$}}}%
    \put(0.9210104,0.46225031){\makebox(0,0)[lb]{\smash{$x^1$}}}%
    \put(0.8318154,0.52253869){\makebox(0,0)[lb]{\smash{$\phi$}}}%
  \end{picture}%
\endgroup
\end{center}
\caption{The 2d lattice of a plane transverse to the defect line. The projection  of the 
defect plane is the heavier line, crossed by frustrated links, which plays the r\^ole of a $\mathbb{Z}_2$
monodromy cut.}
\label{fig:2d_4spins}
\end{figure}
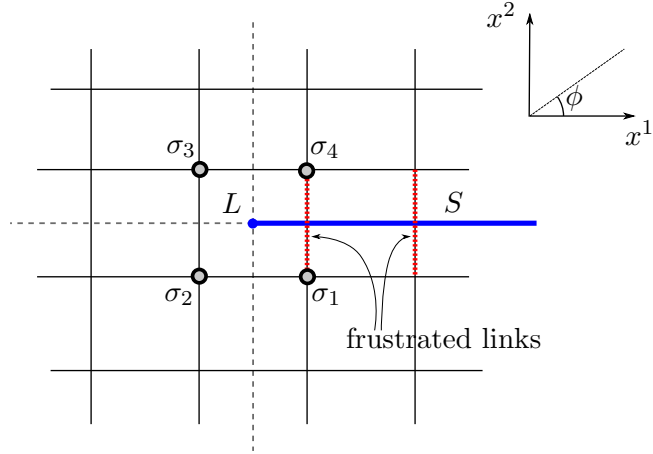
Let us thus consider the symmetry of the lattice plane orthogonal to the 
defect line, as in Fig.s \ref{fig:setup}, \ref{fig:2d_4spins}.
In absence of the monodromy cut corresponding to the projection of the defect $S$, the symmetry of 
the square lattice would be given by the $D_4$ dihedral group, generated by the rotation of $\pi/2$ (say, counter-clockwise) and the reflection about any of the symmetry axes (say, the horizontal one in Fig. \ref{fig:2d_4spins}). In presence of the cut, the counter-clockwise rotation of $\pi/2$ must be accompanied by a ``gauge'' transformation to bring the defect to its original position; in doing so, the spin variables which switch side w.r.t. the frustrated plane change sign; 
we denote this transformation as $a$. Its action on the elementary spins $\sigma_i$ ($i=1,2,3,4$) at 
the corners of a plaquette ``linked'' with 
the defect (as depicted in Fig. \ref{fig:2d_4spins}) is as follows:
\begin{equation}
\label{aact}
a\,:~ (\sigma_1,\sigma_2,\sigma_3,\sigma_4)\mapsto 
(\sigma_2,\sigma_3,\sigma_4,-\sigma_1)~. 
\end{equation}
Indeed, bringing the cut to its original position after the rotation, it crosses the $\sigma_1$ spin variable, switching its sign.

The reflection $b$ with respect to the axis through the origin containing the 
projection of the frustrated plane (the $x^1$ axis) is not affected by the presence of the cut and acts on the spins $\sigma_i$ as follows:
\begin{equation}
\label{bact}
b\,:~ (\sigma_1,\sigma_2,\sigma_3,\sigma_4)\mapsto 
(\sigma_4,\sigma_3,\sigma_2,\sigma_1)~. 
\end{equation}

These transformations satisfy $a^8=\mathbb{1}$, $b^2=\mathbb{1}$ and $(ab)^2=\mathbb{1}$, and generate thus the dihedral group $D_8$. Thus, in our discrete lattice set-up, the symmetry group in the plane is effectively augmented in presence of the monodromy, and mixes the space-time and the flavor $\mathbb{Z}_2$ symmetry; Coleman-Mandula theorem does not apply in this case.

Besides the $D_8$ invariance, there is another symmetry of the theory which turns out to be useful in the classification of the local operators on the defect, namely the reflection with respect to  a plane orthogonal to the defect line; if $x^3$ is the coordinate along the defect, the reflection with respect to such a plane through the origin is of course
\begin{equation}
\mathcal{S}\,:~x^3\mapsto -x^3~,
\label{sparity}
\end{equation}
and we call the parity of an operator with respect this symmetry $\mathcal{S}$-parity.

The $D_8$ group has order 16 and possesses seven irrepses, four of dimension 1 and three of dimension 2. The 4-dimensional representation acting on the 
spins $\sigma_i$ of Fig. \ref{fig:2d_4spins}
decomposes into two bi-dimensional representations, which we denote as $H_{1/2}$ and $H_{3/2}$.
A basis for $H_{1/2}$ is given by $(\psi,\psi^*)$, with
\begin{equation}
\label{defh1}
\psi = \sigma_1+\omega \sigma_2+\omega^2 \sigma_3+\omega^3 \sigma_4\,,
\end{equation}
where $\omega=\exp(\ii\pi/4)$.
In this representation we have 
 \begin{equation}
a\left(\begin{matrix}
\psi\cr
\psi^*\cr
\end{matrix}\right) =\left(
\begin{matrix}
\omega^{-1}&0\cr
0&\omega\cr
\end{matrix}
\right)\, 
\left(
\begin{matrix}
\psi\cr
\psi^*\cr
\end{matrix} 
\right)\,,~~
b\left(\begin{matrix}
\psi\cr
\psi^*\cr
\end{matrix}\right) =\left(
\begin{matrix}
0&\omega^{3}\cr\omega^{-3}
&0\cr
\end{matrix}
\right)\, 
\left(
\begin{matrix}
\psi\cr
\psi^*\cr
\end{matrix} 
\right)~.
\end{equation}

With our conventions, a field $\varphi$ transforming under $a$  as
 $a\,\varphi=\ee^{-J\ii\pi/2}\varphi$ has spin $J$, so $\psi$ ($\psi^*$) carries spin 
$J=1/2$ ($J=-1/2$). 

The basis for the representation $H_{3/2}$ is instead given by
$(\psi_{3/2},\psi_{3/2}^*)$, with    
\begin{equation}
\label{defh2}
\psi_{3/2}= \sigma_1+\omega^3 \sigma_2+\omega^6 \sigma_3+\omega \sigma_4\,.
\end{equation}
In this representation we have
\begin{equation}
a\left(\begin{matrix}
\psi_{3/2}\cr
\psi_{3/2}^*\cr
\end{matrix}\right) =
\left(
\begin{matrix}
\omega^{-3}&0\cr
0&\omega^{3}\cr
\end{matrix}
\right)\, 
\left(
\begin{matrix}
\psi_{3/2}\cr
\psi_{3/2}^*\cr
\end{matrix} 
\right)\,,~~
b\left(\begin{matrix}
\psi_{3/2}\cr
\psi_{3/2}^*\cr
\end{matrix}\right) =\left(
\begin{matrix}
0&\omega\cr
\omega^{-1}
&0\cr
\end{matrix}
\right)\, 
\left(
\begin{matrix}
\psi_{3/2}\cr
\psi_{3/2}^*\cr
\end{matrix} 
\right)~.
\end{equation}

Both the representations $H_{1/2}$ and $H_{3/2}$ are odd under the the flavor group $\mathbb{Z}_2$. All  other irreducible representations of $D_8$ are  $\mathbb{Z}_2$ even and can be obtained by decomposing the direct product of $H_{1/2}$'s. Three of them can be realized in terms of the bilinears $\sigma_i \sigma_j$
corresponding to the four links $\vev{ij}$ of Fig. \ref{fig:2d_4spins} (see also Fig. \ref{fig:bozzi}).
Indeed the 4-dimensional representation $\cL_4$ acting on the four links can be decomposed as the sum of 
a bi-dimensional representation, which we call $V$, and two unidimensional representations:
$\cL_4=V\oplus S\oplus T_+$. Here $S$ is the trivial representation, which acts on
\begin{equation}
\label{defS}
s=\sigma_1 \sigma_2+ \sigma_2 \sigma_3 + \sigma_3 \sigma_4 - \sigma_4 \sigma_1
\end{equation}
by $a\,s=s$, $b\,s=s$. The basis elements $(D_1,D_2)$ of the two-dimensional vectorial representation $V$, corresponding to spin $J=1$, can be chosen to be
\begin{equation}
\label{defV}
D_1= \sigma_1 \sigma_2 + \sigma_2 \sigma_3 - \sigma_3 \sigma_4 + \sigma_4 \sigma_1\,,~~~ 
D_2=- \sigma_1 \sigma_2 + \sigma_2 \sigma_3 + \sigma_3 \sigma_4 + \sigma_4 \sigma_1\,.
\end{equation}
In this representation the generators act as follows:
\begin{equation}
a\left(\begin{matrix}
D_1\cr
D_2\cr
\end{matrix}\right) =
\left(
\begin{matrix}
0&1\cr
-1&0\cr
\end{matrix}
\right)\, 
\left(
\begin{matrix}
D_1\cr
D_2\cr
\end{matrix} 
\right)\,,~~
b\left(\begin{matrix}
D_1\cr
D_2\cr
\end{matrix}\right) =\left(
\begin{matrix}
0&1\cr
1&0\cr
\end{matrix}
\right)\, 
\left(
\begin{matrix}
D_1\cr
D_2\cr
\end{matrix} 
\right)~.
\label{tranD}
\end{equation} 
Notice that the generator $a$ acts on the the combination $D=D_1 + \ii D_2$ as $a\, D = \ee^{-\ii\pi/2}\,D$,
so that $D$ has spin $1$, while, of course, $\bar D=D_1 - \ii D_2$ has spin $-1$.
Finally, $T_+$ is a representation of spin $J=2$ acting on
\begin{equation}
\label{defT+}
t_+= \sigma_1 \sigma_2 - \sigma_2 \sigma_3 + \sigma_3 \sigma_4 + \sigma_4 \sigma_1,
\end{equation}
by $a\,t_+=-t_+$, $b\,t_+=t_+$ 

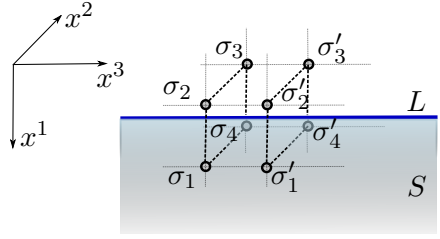
\begin{figure}
\begin{center}

\begingroup
  \makeatletter
  \providecommand\color[2][]{%
    \errmessage{(Inkscape) Color is used for the text in Inkscape, but the package 'color.sty' is not loaded}
    \renewcommand\color[2][]{}%
  }
  \providecommand\transparent[1]{%
    \errmessage{(Inkscape) Transparency is used (non-zero) for the text in Inkscape, but the package 'transparent.sty' is not loaded}
    \renewcommand\transparent[1]{}%
  }
  \providecommand\rotatebox[2]{#2}
  \ifx\svgwidth\undefined
    \setlength{\unitlength}{160pt}
  \else
    \setlength{\unitlength}{\svgwidth}
  \fi
  \global\let\svgwidth\undefined
  \makeatother
  \begin{picture}(1,0.51757199)%
    \put(0,0){\includegraphics[width=\unitlength]{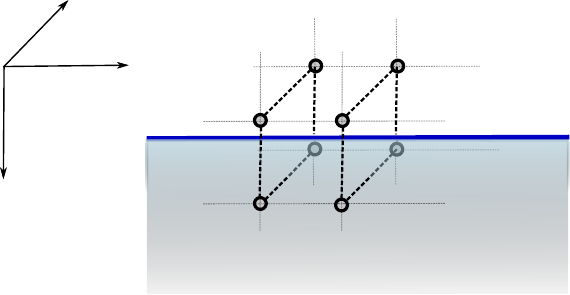}}%
    \put(0.36848712,0.1198131){\makebox(0,0)[lb]{\smash{$\sigma_1$}}}%
    \put(0.35707906,0.33094704){\makebox(0,0)[lb]{\smash{$\sigma_2$}}}%
    \put(0.47955993,0.42681225){\makebox(0,0)[lb]{\smash{$\sigma_3$}}}%
    \put(0.46703896,0.23049546){\makebox(0,0)[lb]{\smash{$\sigma_4$}}}%
    \put(0.92855886,0.09116083){\makebox(0,0)[lb]{\smash{$S$}}}%
    \put(0.92704694,0.29164651){\makebox(0,0)[lb]{\smash{$L$}}}%
    \put(0.20180625,0.34803355){\makebox(0,0)[lb]{\smash{$x^3$}}}%
    \put(0.12763782,0.48924939){\makebox(0,0)[lb]{\smash{$x^2$}}}%
    \put(0.02329965,0.2094854){\makebox(0,0)[lb]{\smash{$x^1$}}}%
    \put(0.60971153,0.11940285){\makebox(0,0)[lb]{\smash{$\sigma_1'$}}}%
    \put(0.63397643,0.31513257){\makebox(0,0)[lb]{\smash{$\sigma_2'$}}}%
    \put(0.71916278,0.423159){\makebox(0,0)[lb]{\smash{$\sigma_3'$}}}%
    \put(0.7074526,0.22197775){\makebox(0,0)[lb]{\smash{$\sigma_4'$}}}%
  \end{picture}%
\endgroup
\end{center}
\caption{The spin variables around the monodromy line in terms of which the defect operators of lowest dimensions
of table \ref{tab:anomalous} can be realized, as described in the text.}
\label{fig:spincubo}
\end{figure}

The $D_8$ representations constructed up to now are defined on a single plane orthogonal to the defect line, thus are  all even under $\mathcal{S}$-parity
(\ref{sparity}).
 There are two more unidimensional representations of $D_8$, which we 
shall denote as $P$ and $T_-$.
$P$ is generated by antisymmetric products of two representations of 
$H_{1/2}$ type, and cannot thus be realized 
in terms of the links $\sigma_i \sigma_j$ of the single plaquette in 
Fig \ref{fig:2d_4spins}.
Their minimal lattice realization involves the cube depicted in Fig. \ref{fig:spincubo}, obtained by adjoining to
the square of Fig. \ref{fig:2d_4spins} its translation of one lattice spacing along the defect line $L$. We denote the spins of this new square with $\sigma_i'$. 
Note that the $\mathcal{S}$-reflection defined in (\ref{sparity}) now is
\begin{equation}
\mathcal{S}\,:\sigma_i\leftrightarrow\sigma_i'~.
\end{equation}
\dgb
\begin{table}[ht]
\begin{center}
\begin{tabular}{|cccc|}
\hline
$s$&$p^o$&$D$&$D^o$\\
\hline
$\bar{\psi}\psi$&$\Im m(\bar{\psi}\vp\psi)$&
$\ii\psi\psi$&$\bar{\psi}\vp\psi_{3/2}$
\\
\hline
\end{tabular}
\end{center}
\begin{center}
\begin{tabular}{|cccc|}
\hline
$t_+$&$t_-$&$t^o_+$&$t^o_-$\\
\hline
$\Im m(\psi\psi_{3/2})$&$\Re e(\psi\psi_{3/2})$&$\Im m(\psi\vp\psi_{3/2})$&
$\Re e(\psi\vp\psi_{3/2})$\\
\hline
\end{tabular}
\end{center}
\caption{Schematic description of several lattice operators, built as bilinears in $\psi$ and 
$\psi_{3/2}$ in analogy to the free field approximation to the primary operators of the continuum theory. 
$\partial f$ denotes the finite difference $\partial f(x)\equiv f(x+1)-f(x)$ and $g\vp f=g\partial f-f\partial g$.
Using the transformation properties of $\psi$ and $\psi_{3/2}$ one can verify at once the transformation properties of these bilinears, in accordance with the decomposition of representations described in \eq{decompD8} and Fig. \ref{fig:D8}. As described in the text, some of these operators can be built explicitly from the spins at the vertices of a single plaquette, 
some require us to use spins from the vertices of a cube. These lattice operators provide natural candidates for the corresponding operators in the continuum theory, 
up to some ambiguity due to the fact that the spin $J$ on the lattice is defined modulo $4$.}
\label{tab:bilinears} 
\end{table}
\dge
The anti-symmetric combinations $\cD_4$ of the diagonals of the four faces which do not intersect the defect line  define a four-dimensional 
reducible representation of $D_8$ which can be decomposed as 
$\cD_4=P\oplus V\oplus T_-$, where now the representations $P,V$ and $T_-$ act on $\mathcal{S}$-odd operators, that we call respectively $p^o$,$D^o$, 
and $t_-^o$. 
We have
\beqa
\label{po}
p^o=&[\sigma_1\sigma_2']+[\sigma_2\sigma_3']+[\sigma_3\sigma_4']-
[\sigma_4\sigma_1']~,\\
D^o_1=&-[\sigma_1\sigma_2']+[\sigma_2\sigma_3']+[\sigma_3\sigma_4']+
[\sigma_4\sigma_1']~,\,\\
D^o_2=&[\sigma_1\sigma_2']+[\sigma_2\sigma_3']-[\sigma_3\sigma_4']+
[\sigma_4\sigma_1']~,\\
t^o_-=&[\sigma_1\sigma_2']-[\sigma_2\sigma_3']+[\sigma_3\sigma_4']+
[\sigma_4\sigma_1']~,
\eeqa 
with $[\sigma_i\sigma_j']= \sigma_i\sigma_j' -\sigma_i'\sigma_j$.
The transformation of $D^o$ under $a$ and $b$ is the same 
of that of $D$, defined in (\ref{tranD}), while for the other two operators we have 
$a\,p^o=p^o$, $b\,p^o=-p^o$ and $a\,t^o_-=-t^o_-$, $b\,t^o_-=-t^o_-$.
Similarly the anti-symmetric combinations $\cD_2$ of the principal 
diagonals of this cube can be decomposed in the sum $\cD_2= T_+\oplus P$.  
The pseudoscalar representation $T_+$ now acts on
\begin{equation}
t_+^o= [\sigma_1 \sigma_3']-  [ \sigma_2 \sigma_4']
\end{equation}
as in (\ref{defT+}).
The pseudoscalar representation $P$, instead,  acts now on the $\mathcal{S}$-odd operator
\begin{equation}
p^{o'}= [\sigma_1 \sigma_3'] + [\sigma_2 \sigma_4']~,
\end{equation}
which is however less efficient in numerical simulations than $p^o$ introduced in 
\eq{po}. 

 It is not difficult to convince oneself that for any irreducible 
representation of $D_8$ one can construct primary operators of both 
$\mathcal{S}$-parities. Those described so far can be also expressed as bilinears 
in $\psi$ and/or $\psi_{3/2}$ as shown in tab. \ref{tab:bilinears}.
The other primary operators of opposite $\mathcal{S}$-parity can be expressed as 
multi-linear products of $\sigma_i$ or as operators involving more nodes of the lattice; they are thus difficult to deal with in numerical simulations and are 
expected to have larger anomalous dimensions.    

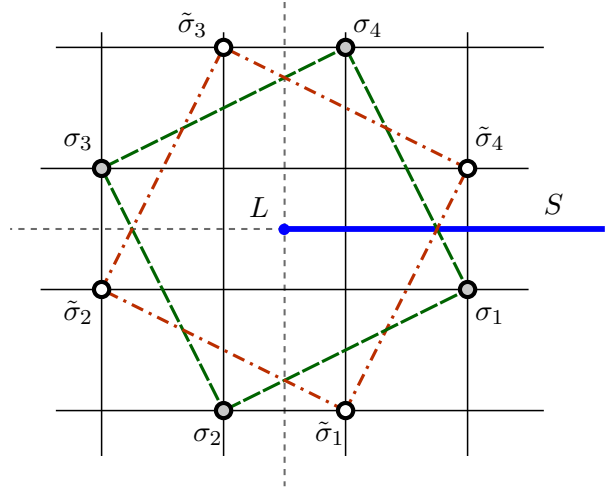
\begin{figure}
\begin{center}

\begingroup
  \makeatletter
  \providecommand\color[2][]{%
    \errmessage{(Inkscape) Color is used for the text in Inkscape, but the package 'color.sty' is not loaded}
    \renewcommand\color[2][]{}%
  }
  \providecommand\transparent[1]{%
    \errmessage{(Inkscape) Transparency is used (non-zero) for the text in Inkscape, but the package 'transparent.sty' is not loaded}
    \renewcommand\transparent[1]{}%
  }
  \providecommand\rotatebox[2]{#2}
  \ifx\svgwidth\undefined
    \setlength{\unitlength}{250pt}
  \else
    \setlength{\unitlength}{\svgwidth}
  \fi
  \global\let\svgwidth\undefined
  \makeatother
  \begin{picture}(1,0.73071906)%
    \put(0,0){\includegraphics[width=\unitlength]{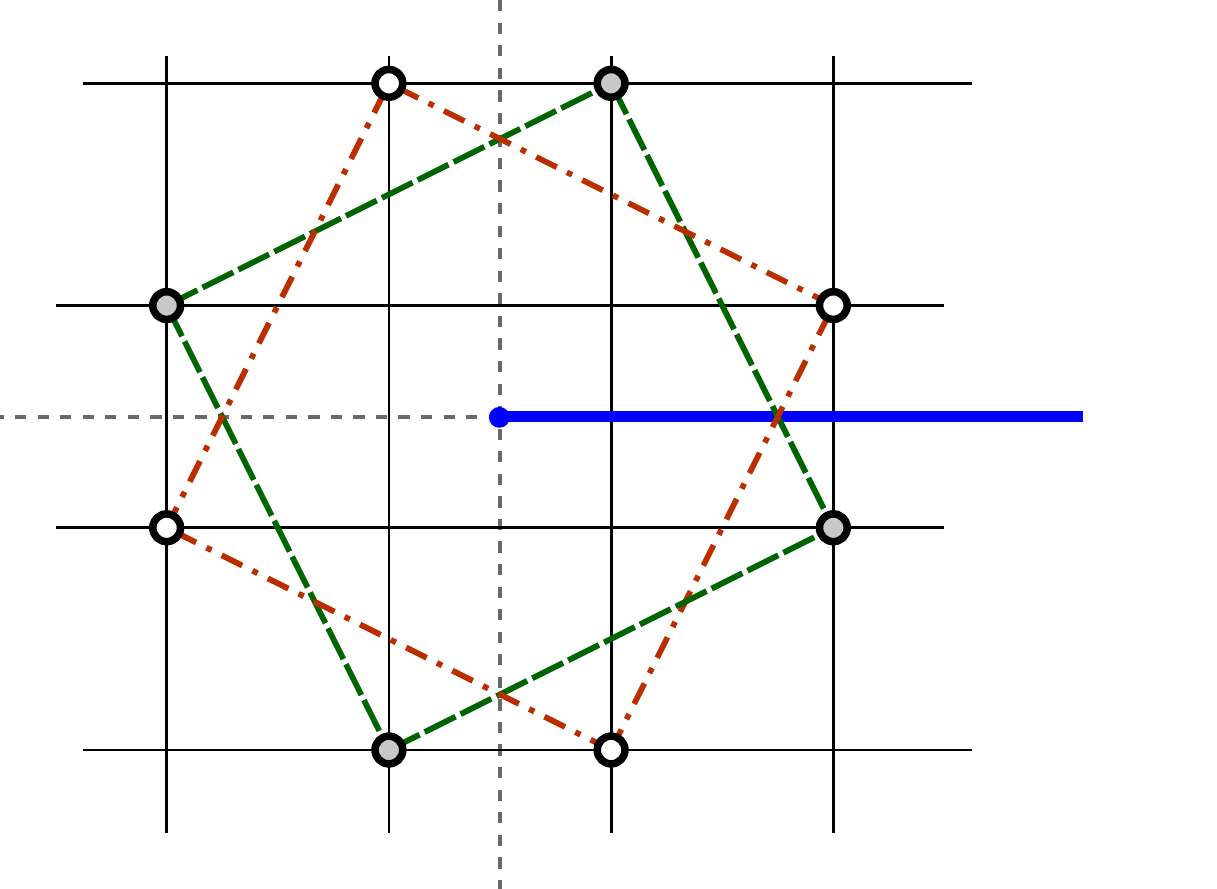}}%
    \put(0.6964666,0.251183){\makebox(0,0)[lb]{\smash{$\sigma_1$}}}%
    \put(0.51378684,0.68504744){\makebox(0,0)[lb]{\smash{$\sigma_4$}}}%
    \put(0.0799224,0.51378516){\makebox(0,0)[lb]{\smash{$\sigma_3$}}}%
    \put(0.27401965,0.06850324){\makebox(0,0)[lb]{\smash{$\sigma_2$}}}%
    \put(0.79922397,0.4110278){\makebox(0,0)[lb]{\smash{$S$}}}%
    \put(0.35819674,0.40587118){\makebox(0,0)[lb]{\smash{$L$}}}%
    \put(0.6964666,0.51378516){\makebox(0,0)[lb]{\smash{${\tilde\sigma}_4$}}}%
    \put(0.25118468,0.68504744){\makebox(0,0)[lb]{\smash{${\tilde\sigma}_3$}}}%
    \put(0.45669941,0.06850324){\makebox(0,0)[lb]{\smash{${\tilde\sigma}_1$}}}%
    \put(0.0799224,0.251183){\makebox(0,0)[lb]{\smash{${\tilde\sigma}_2$}}}%
  \end{picture}%
\endgroup
\end{center}
\caption{An alternative set of spin variables  in terms of which we construct some of the defect operators, as described in the text. These spins are lying around the monodromy line in plane orthogonal to it, just as in Fig. \ref{fig:2d_4spins}. In particular, the \dgb $\mathcal{S}$-even \dge pseudo-scalar representation $P$ can be realized in terms of the bilinears $\sigma_i\sigma_{i+1}$ and $\tilde\sigma_i\tilde\sigma_{i+1}$, indicated in the drawing by the two sets of diagonal segments.}
\label{fig:altspins}
\end{figure}

As an example we describe  a simple realization 
of an $\mathcal{S}$-even pseudoscalar $\tilde{p}$ which can  be obtained by 
considering, instead of the eight spin variables on the cube of 
Fig. \ref{fig:spincubo}, eight spin variables lying around the line 
defect as in Fig. \ref{fig:altspins}.
The generators act on these variables as follows:
\begin{equation}
\label{geneight}
\begin{aligned}
a\,:~ &  (\sigma_1,\sigma_2,\sigma_3,\sigma_4)\mapsto (\sigma_2,\sigma_3,\sigma_4,-\sigma_1)~,~~
(\tilde\sigma_1,\tilde\sigma_2,\tilde\sigma_3,\tilde\sigma_4)\mapsto 
(\tilde\sigma_2,\tilde\sigma_3,\tilde\sigma_4,-\tilde\sigma_1)~;\\
b\,:~ & \sigma_1\leftrightarrow \tilde\sigma_4~,~~ \sigma_4\leftrightarrow \tilde\sigma_1~,~~
 \sigma_2\leftrightarrow \tilde\sigma_3~,~~ \sigma_3\leftrightarrow \tilde\sigma_2~. 
\end{aligned}
\end{equation}
It is easy to check that the $\mathbb{Z}_2$-even operator 
\begin{equation}
\label{defpp}
\tilde{p} = \sigma_1\sigma_2 + \sigma_2\sigma_3 + \sigma_3\sigma_4 - \sigma_1\sigma_4 -
\left(\tilde\sigma_1\tilde\sigma_2 + \tilde\sigma_2\tilde\sigma_3 + \tilde\sigma_3\tilde\sigma_4 - \tilde\sigma_1\tilde\sigma_4\right)
\end{equation}
transforms in the pseudoscalar representation $P$, i.e., 
we have $a\tilde p = \tilde p$ and $b\tilde p = -\tilde p$.

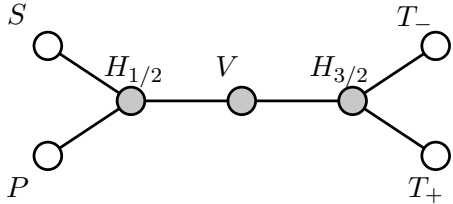
\begin{figure}
\begin{center}

\begingroup
  \makeatletter
  \providecommand\color[2][]{%
    \errmessage{(Inkscape) Color is used for the text in Inkscape, but the package 'color.sty' is not loaded}
    \renewcommand\color[2][]{}%
  }
  \providecommand\transparent[1]{%
    \errmessage{(Inkscape) Transparency is used (non-zero) for the text in Inkscape, but the package 'transparent.sty' is not loaded}
    \renewcommand\transparent[1]{}%
  }
  \providecommand\rotatebox[2]{#2}
  \ifx\svgwidth\undefined
    \setlength{\unitlength}{180pt}
  \else
    \setlength{\unitlength}{\svgwidth}
  \fi
  \global\let\svgwidth\undefined
  \makeatother
  \begin{picture}(1,0.42040475)%
    \put(0,0){\includegraphics[width=\unitlength]{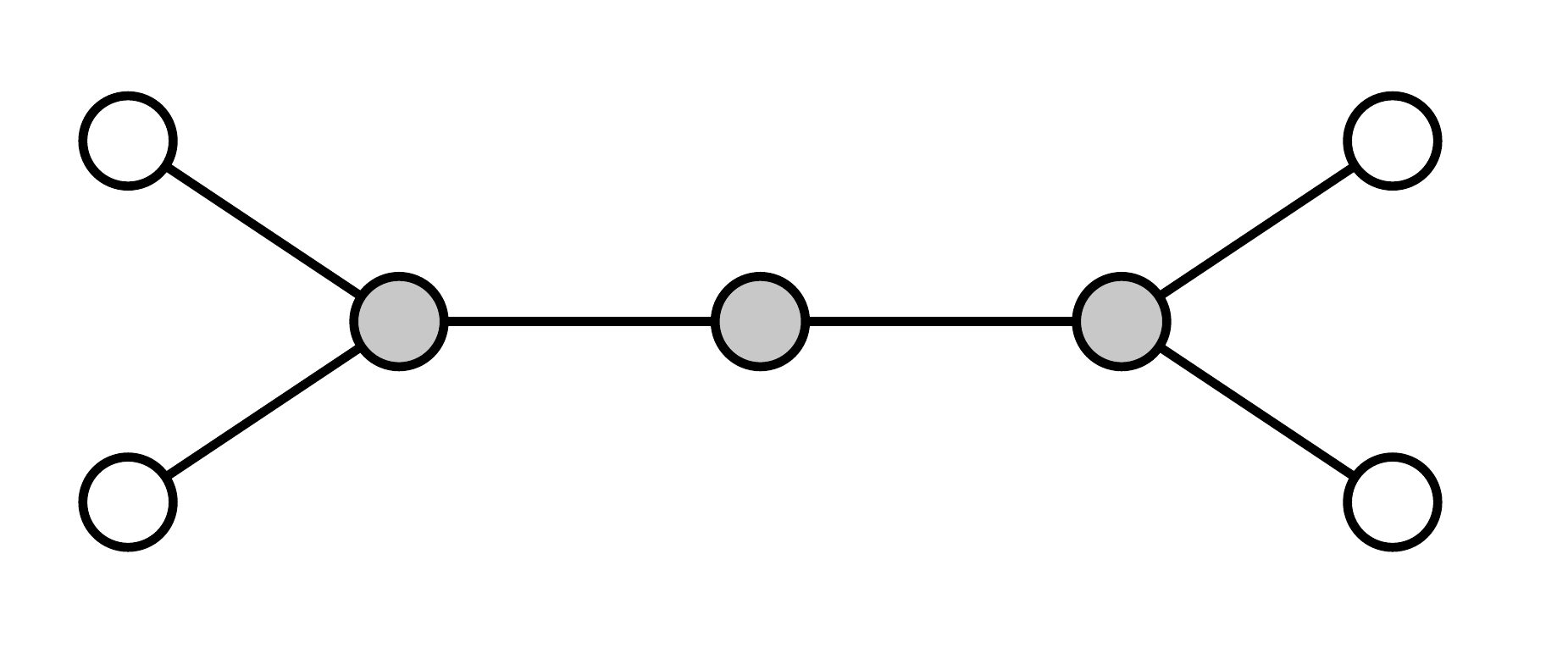}}%
    \put(-0.00286717,0.37662679){\makebox(0,0)[lb]{\smash{$S$}}}%
    \put(-0.00478294,0.01358917){\makebox(0,0)[lb]{\smash{$P$}}}%
    \put(0.8307121,0.01358917){\makebox(0,0)[lb]{\smash{$T_+$}}}%
    \put(0.80764927,0.37279525){\makebox(0,0)[lb]{\smash{$T_-$}}}%
    \put(0.19785668,0.26704362){\makebox(0,0)[lb]{\smash{$H_{1/2}$}}}%
    \put(0.43051699,0.26704362){\makebox(0,0)[lb]{\smash{$V$}}}%
    \put(0.62565144,0.26704362){\makebox(0,0)[lb]{\smash{$H_{3/2}$}}}%
  \end{picture}%
\endgroup
\end{center}
\caption{The irrepses of the D8 dihedral group are encoded in the extended Dynkin diagram of the $\mathcal{D}_6$ algebra, as described in the text. 
Open circles denote one-dimensional representations while grey circles are associated to two-dimensional representations.}
\label{fig:D8}
\end{figure}

An useful tool to summarize the $D_8$ irrepses we discussed above is the graph associated with the decomposition of the tensor product of any irreducible representation $R_i$ with the two-dimensional representation $H_{1/2}$, 
see Fig. \ref{fig:D8}. 
The incidence matrix of this graph%
\footnote{This is very similar to the McKay correspondence between discrete SU$(2)$ subgroups and extended Dynkin diagrams of ADE type.}, which turns out to correspond to the extended Dynkin diagram of the $\cD_6$ Lie 
Algebra, is given by  the Clebsh-Gordan coefficients $c_{ij}$ in the decomposition
\begin{equation}
\label{decompD8}
 H_{1/2} \otimes R_i =\sum_{j}c_{ij} R_j~.
\end{equation}

\begin{table}[ht]
\centering
\begin{tabular}{|cccccc|}
\hline
 $\mathfrak{o}$&D8 irrep  & $ \Z_2 $ parity & $ O(2) $ spin &$\mathcal{S}$-parity& $ \Delta $ \\
\hline
 $ s $ &$S$  & + & $ 0^+ $ &+& 2.27(1)  \\
 $ p^o $ &$P$ & + & $ 0^- $ &--& 2.9(2) \\
 $ \tilde{p} $ &$P$ & + & $ 0^- $&+& 3.7(2)[3] \\
 $ \psi$ &$H_{1/2}$ & -- & $ \frac{1}{2} $&+  & 0.9187(6) \\
 $ D $&$V$  & + & $  1  $&+ & 2 \\
$ D^o$&$V$  & + & $  1  $&-- & 3.3(2)[3] \\
 $ \psi_{3/2}$&$H_{3/2}$ & -- & $ \frac{3}{2} $&+  & 1.99(5) \\
 $ t_+ $&$T_+$ & + & $ 2 $&+ & 3.1(5)[3] \\
 $ t_+^o,t_-^o $&$T_+\,,T_-$ & + & $ 2 $&-- &  $\ge$4.2(1) \\
\hline
\end{tabular}
\caption{The lowest anomalous dimensions of the local operators
at the defect line. The round brackets indicate the statistical error, while 
the square brackets for $\Delta\ge3$ denote an estimate of the 
systematic error (see more details in the text). For the operator  $t_+^o$, 
realized on the diagonal of elementary cubes, we did not use a 
direct Monte Carlo evaluation which is too noisy, but a lower bound 
obtained by considering only the contributions of the spin-spin 
two-point functions involved.}
\label{tab:anomalous}
\end{table}

\section{Results}
\label{Section:Results}
We have seen in section \ref{representations} that most of $D_8$ representations can be realized using the links and the nodes of a plaquette topologically linked with the defect as shown in fig. \ref{fig:2d_4spins}. 

\begin{figure}
\includegraphics[width=14 cm]{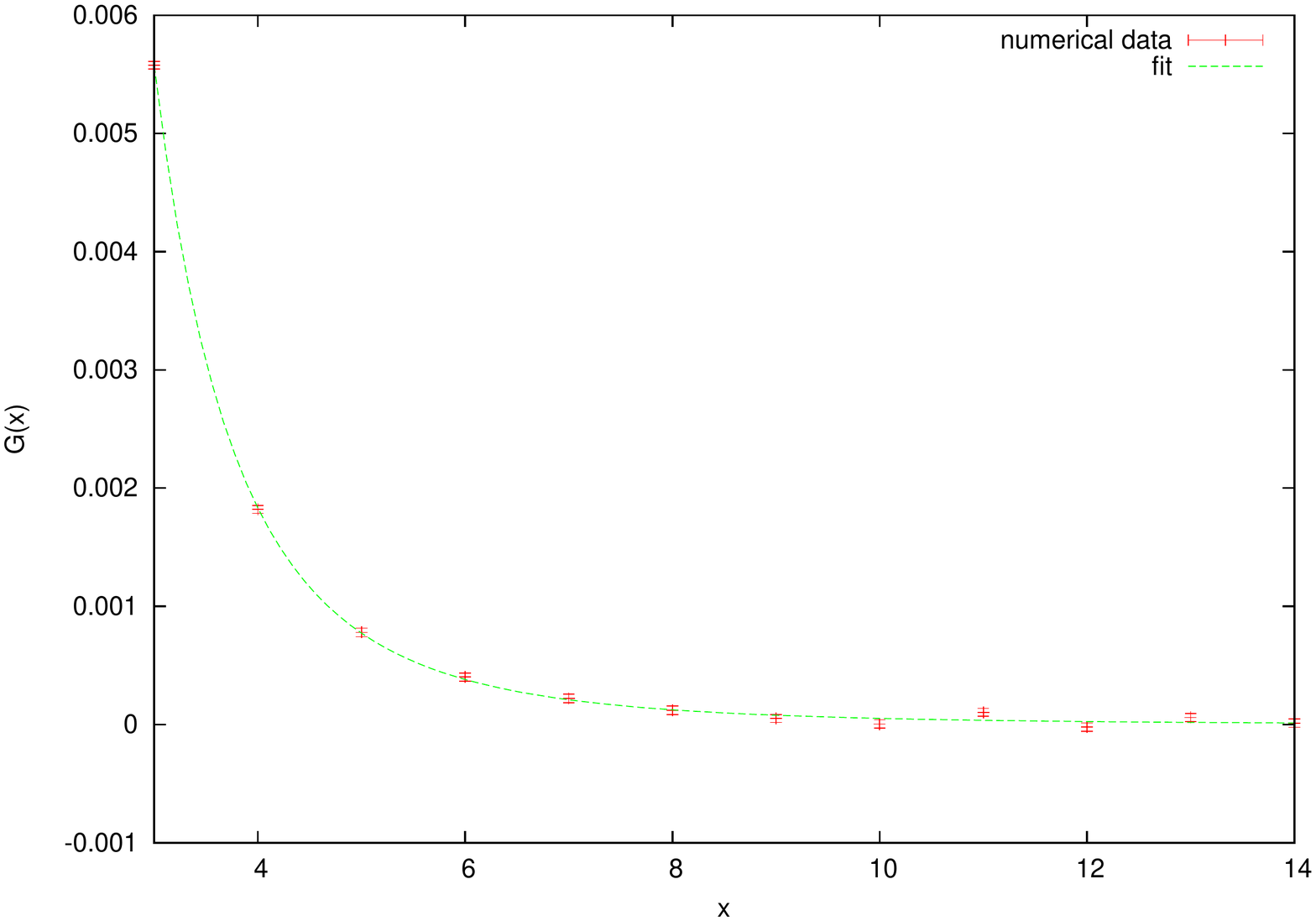}
\caption{The correlation function of the displacement operator. 
The solid curve is the one-parameter fit to $a_{DD}/x^4$.}
\label{fig:displacement}
\end{figure}

In a first series of numerical experiments we evaluated the correlation function along the defect line, between a pair of links of this type located  at 
a mutual distance $x$. We performed an independent simulation for every value of $x$ and for every  orientation of the pair of links. We then arranged these correlation functions in irreducible representations of $D_8$, according to the prescriptions of Eq.s (\ref{defS}, \ref{defV}, \ref{defT+}) 
for the operators  $s$, $D$ and $t_+$,
in order to extract the exponent of their power-like decay. We expect, for the correlators
on the defect of such operators, the behavior
\begin{equation}
\vev{s(0)s(x)}=a_{ss}\,x^{-2\Delta_s}+\mathrm{constant}\,,~~~
\vev{\bar{D}(0)D(x)}=a_{DD}\,x^{-2\Delta_D}\,,~~~
\vev{t_+(0)t_+(x)}=a_{tt}\,x^{-2\Delta_t}\,.
\label{twopoint}
\end{equation}
We checked in all cases this power behavior, even if the 
evaluation of the exponents $\Delta$ cannot be very accurate,
due to the fact it turns out that $\Delta\geq 2$.
This implies that the correlation function falls off 
very rapidly and  after few lattice spacings the signal is drowned in noise 
(this tendency is already evident for the correlator of the operator $D$, that has $\Delta_D=2$,
depicted in Fig. \ref{fig:displacement}, even if in this case the fit is still very good). When the exponents are $\Delta\geq 3$ the accuracy of their determination becomes problematic because the value depends on the way we fit the data. 
This means that these evaluations are affected by a systematic error.
In order to get an idea of the size of this error, we fitted the data with two different procedures, namely to a single power law, like in (\ref{twopoint}), or to the binomial $a/x^{2\Delta}+b/x^{2(\Delta+1)}$ taking into account also 
the contribution of the first secondary operator which can contribute; we used the difference 
between these two determinations of $\Delta$ as a rough estimate of the 
systematic error, which is reported in square brackets in Tab. 
\ref{tab:anomalous}.

\dgb At small values of the spin $J$, the identification between lattice operators in a given $D_8$ representations and continuum operators in appropriate $O(2)$ representations is clear. 
The operators with high spin in the continuum theory are expected to have large conformal dimension, and thus give subleading contributions to the lattice operators. 
The one exception is the operator $\tilde p$. Although it has no spin from the point of view of $D_8$, it is built from alternating spins in 
a way which would closely resemble a $J=4$ operator of the continuum theory. In the free theory, a $J=0$ operator with the quantum numbers of $\tilde p$ would be a descendant of 
$p^o$. The $J=4$ primary would have dimension close to $5$. As the lattice correlation function of $p^o$ and $\tilde p$ is very small, 
the $J=4$ contribution to $\tilde p$ must be very large, and the numerical estimates correspondingly poor. 
\dge     

In the case of the scalar $s$ we can  conveniently extract 
$\Delta_s$ from the one-point function, which is expected to have the functional form
\begin{equation}
\vev{s}=\frac{a_s}{\ell^{\Delta_s}}+{\rm constant}\,,
\end{equation} 
where $\ell$ is the length of the defect line. Combining this one-point function with the two-point function defined in (\ref{twopoint}) we can also 
extract the universal amplitude ratio $a_s/\sqrt{\vert a_{ss}\vert}$ 
which turns out to be $0.33(1)$.
\begin{figure}
\begin{center}

\begingroup
  \makeatletter
  \providecommand\color[2][]{%
    \errmessage{(Inkscape) Color is used for the text in Inkscape, but the package 'color.sty' is not loaded}
    \renewcommand\color[2][]{}%
  }
  \providecommand\transparent[1]{%
    \errmessage{(Inkscape) Transparency is used (non-zero) for the text in Inkscape, but the package 'transparent.sty' is not loaded}
    \renewcommand\transparent[1]{}%
  }
  \providecommand\rotatebox[2]{#2}
  \ifx\svgwidth\undefined
    \setlength{\unitlength}{420pt}
  \else
    \setlength{\unitlength}{\svgwidth}
  \fi
  \global\let\svgwidth\undefined
  \makeatother
  \begin{picture}(1,0.33938524)%
    \put(0,0){\includegraphics[width=\unitlength]{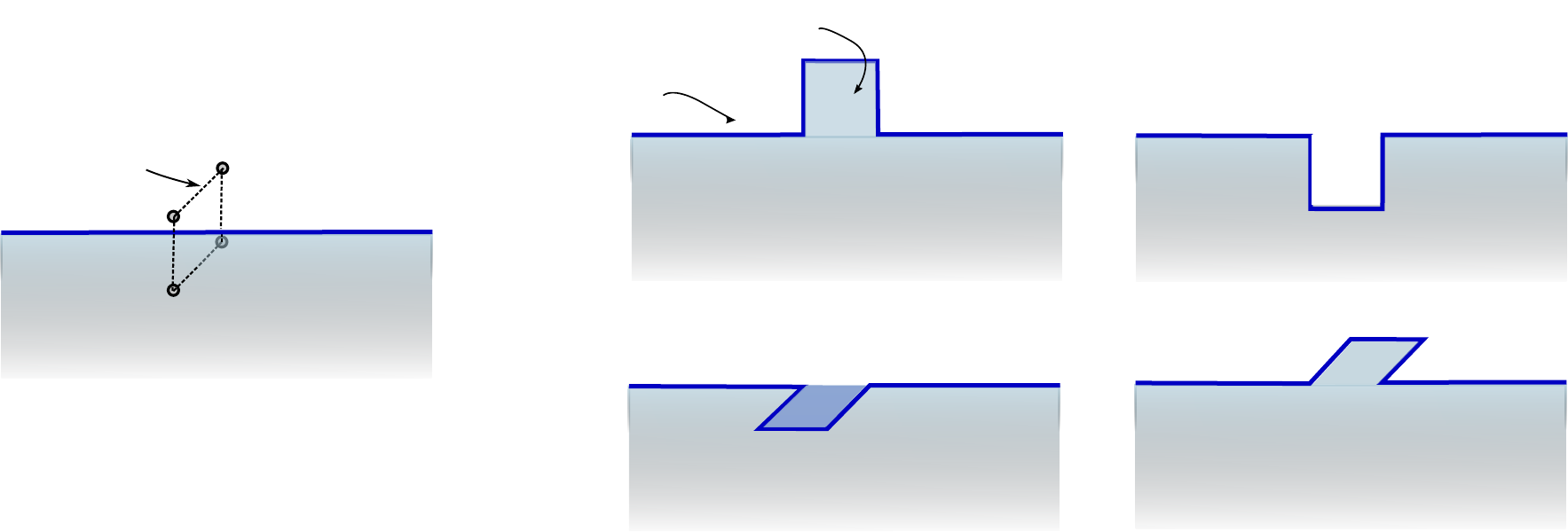}}%
    \put(0.37446723,0.28602251){\makebox(0,0)[lb]{\smash{Wilson line}}}%
    \put(0.46755333,0.32955302){\makebox(0,0)[lb]{\smash{Plaquette dual to $\vev{23}$ }}}%
    \put(0.28401148,0.13267055){\makebox(0,0)[lb]{\smash{$\stackrel{\mathrm{duality}}{\longrightarrow}$}}}%
    \put(0.53971475,0.05998382){\makebox(0,0)[lb]{\smash{$\vev{12}$}}}%
    \put(0.78634291,0.11271236){\makebox(0,0)[lb]{\smash{$\vev{34}$}}}%
    \put(0.83647328,0.22559131){\makebox(0,0)[lb]{\smash{$\vev{41}$}}}%
    \put(0.09145327,0.13690738){\makebox(0,0)[lb]{\smash{$\sigma_1$}}}%
    \put(0.07967354,0.21072001){\makebox(0,0)[lb]{\smash{$\sigma_2$}}}%
    \put(0.12908905,0.24266926){\makebox(0,0)[lb]{\smash{$\sigma_3$}}}%
    \put(0.13942748,0.167973){\makebox(0,0)[lb]{\smash{$\sigma_4$}}}%
    \put(0.05811522,0.23674769){\makebox(0,0)[lb]{\smash{$\vev{23}$}}}%
    \put(0.18604652,0.12583359){\makebox(0,0)[lb]{\smash{$S$}}}%
    \put(0.22890947,0.20005968){\makebox(0,0)[lb]{\smash{$L$}}}%
  \end{picture}%
\endgroup
\end{center}
\caption{Using  Kramers-Wannier duality one can transform the four link 
variables of Fig.\ref{fig:2d_4spins} into staples which deform the 
line defect and can be used to build the displacement operator.}
\label{fig:bozzi}
\end{figure}
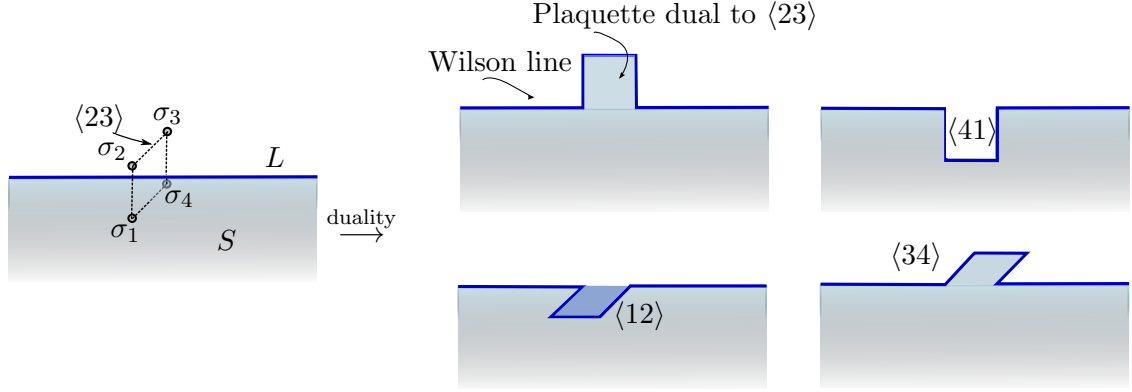

It is worth noting that Kramers-Wannier duality allows to map the 
link variables of the Ising model into the plaquettes of the dual 
$\Z_2$ gauge theory.  Precisely we have, for any coupling $\beta$,
 \begin{equation}
 P_{ij}=\cosh 2\beta - \sigma_i \sigma_j\sinh2\beta\,,
\end{equation}
where $P_{ij}$ is the plaquette of the dual lattice orthogonal to the link 
$\vev{ij}$. Thus correlation functions between  link variables of Fig. 
\ref{fig:2d_4spins} can be written as correlators between staples deforming the 
defect line, as indicated  in Fig. \ref{fig:bozzi}. In particular the 
vector representation $V$ defined in (\ref{defV}) allows to build a 
discretized version of the  displacement operator $D$ discussed in Section 
\ref{Section:ConformalDefects}. This operator has protected quantum numbers and 
in particular has  dimension $\Delta_D=2$. This is nicely confirmed by a 
one-parameter fit of our numerical data, as shown in 
Fig. \ref{fig:displacement}. 

In another set of numerical experiments we evaluated the correlation function between the spin variables associated to the vertices of the plaquette of fig. \ref{fig:2d_4spins}. 
In terms of these we can construct the correlators of the $\Z_2$ odd local operators of semi-integer spin $J=1/2$ and $J=3/2$ defined in (\ref{defh1}) and (\ref{defh2}). 
It turns out that $\Delta_{\psi}$ is slightly less than 1, so that we can follow the signal of 
the correlator for many lattice spacings. The statistical errors are  
so small that the quality of the data cannot be appreciated in a plot; the results are therefore 
explicitly reported in  Table \ref{tab:spinhalf}.
Using these data we can extract a rather precise estimate of the anomalous 
dimensions of this operator, which is given in Table \ref{tab:anomalous} together with 
those of the other lowest-lying operators.  

One emerging feature of the spectrum of anomalous dimensions is that most of the estimated values are  rather close to the values expected in a free-field 
theory. This can be seen by comparing the dimensions of the operators of table \ref{tab:bilinears} obtained in our simulations with the dimensions that the corresponding bilinears would have in a free field theory,
where one would have $[\psi]=1$, $[\psi_{3/2}]=2$ and, of course, $[\partial]=1$.
It turns out that these free-theory values represent in almost all cases the nearest integers to our numerical results.
%
\begin{table}[ht]
\centering
\begin{tabular}{|cc|cc|}
\hline
 $ x $ & $ G_{1/2}(x) $ &$ x $ & $ G_{1/2}(x) $ \\
\hline
   2 & 0.86752(3) & 11 & 0.04529(5) \\
   3 & 0.46136(4) & 12 & 0.03873(5) \\
   4 & 0.28125(4) & 13 & 0.03341(5) \\
   5 & 0.18892(4) & 14 & 0.02925(5) \\
   6 & 0.13592(4) & 15 & 0.02592(4) \\
   7 & 0.10287(4) & 16 & 0.02301(5) \\
   8 & 0.08066(4) & 17 & 0.02071(5) \\
   9 & 0.06518(4) & 18 & 0.01873(5) \\
  10 & 0.05383(4) & 19 & 0.01692(5) \\
 \hline
\end{tabular}
\caption{The values of the spin-$1/2$ correlation function 
$G_{1/2}(x) = \mathrm{Re} \vev{\psi(x)\psi^*(0)}$
on the  defect line.}
\label{tab:spinhalf}
\end{table}

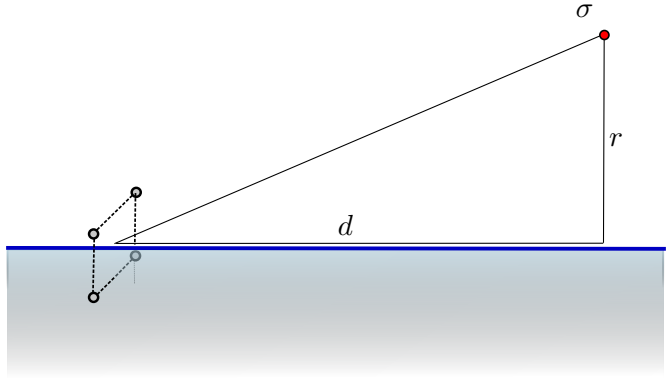
\begin{figure}
\begin{center}

\begingroup
  \makeatletter
  \providecommand\color[2][]{%
    \errmessage{(Inkscape) Color is used for the text in Inkscape, but the package 'color.sty' is not loaded}
    \renewcommand\color[2][]{}%
  }
  \providecommand\transparent[1]{%
    \errmessage{(Inkscape) Transparency is used (non-zero) for the text in Inkscape, but the package 'transparent.sty' is not loaded}
    \renewcommand\transparent[1]{}%
  }
  \providecommand\rotatebox[2]{#2}
  \ifx\svgwidth\undefined
    \setlength{\unitlength}{247pt}
  \else
    \setlength{\unitlength}{\svgwidth}
  \fi
  \global\let\svgwidth\undefined
  \makeatother
  \begin{picture}(1,0.57862595)%
    \put(0,0){\includegraphics[width=\unitlength]{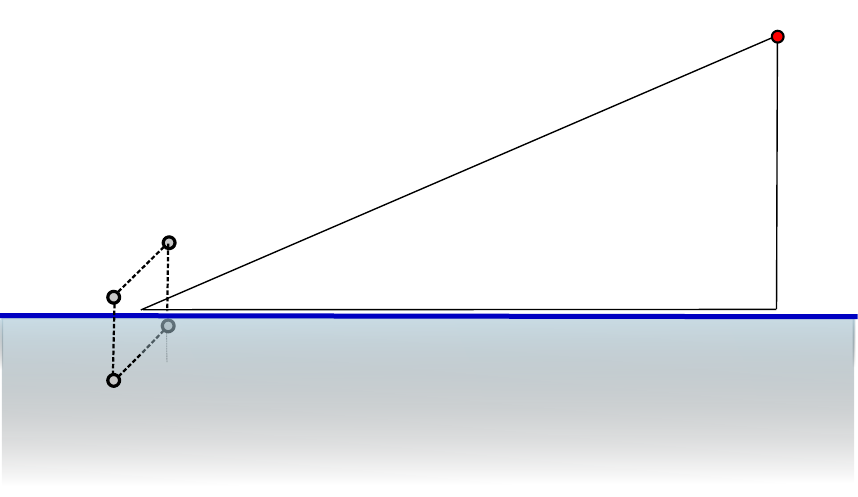}}%
    \put(0.50270022,0.23113337){\makebox(0,0)[lb]{\smash{$d$}}}%
    \put(0.91410896,0.36689206){\makebox(0,0)[lb]{\smash{$r$}}}%
    \put(0.86361326,0.56343427){\makebox(0,0)[lb]{\smash{$\sigma$}}}%
  \end{picture}%
\endgroup
\end{center}
\caption{Set-up for the mixed correlation function of 
the scalar spin on the bulk and the $\Z_2$- odd local operator  
on the defect. This local operator is built with the spins associated 
to the corners of the plaquette wrapped around the defect line, following the prescriptions of Eq.s (\ref{defh1}) and (\ref{defh2}). }
\label{fig:mixed_setup}
\end{figure}

\begin{figure}
\includegraphics[width=14 cm]{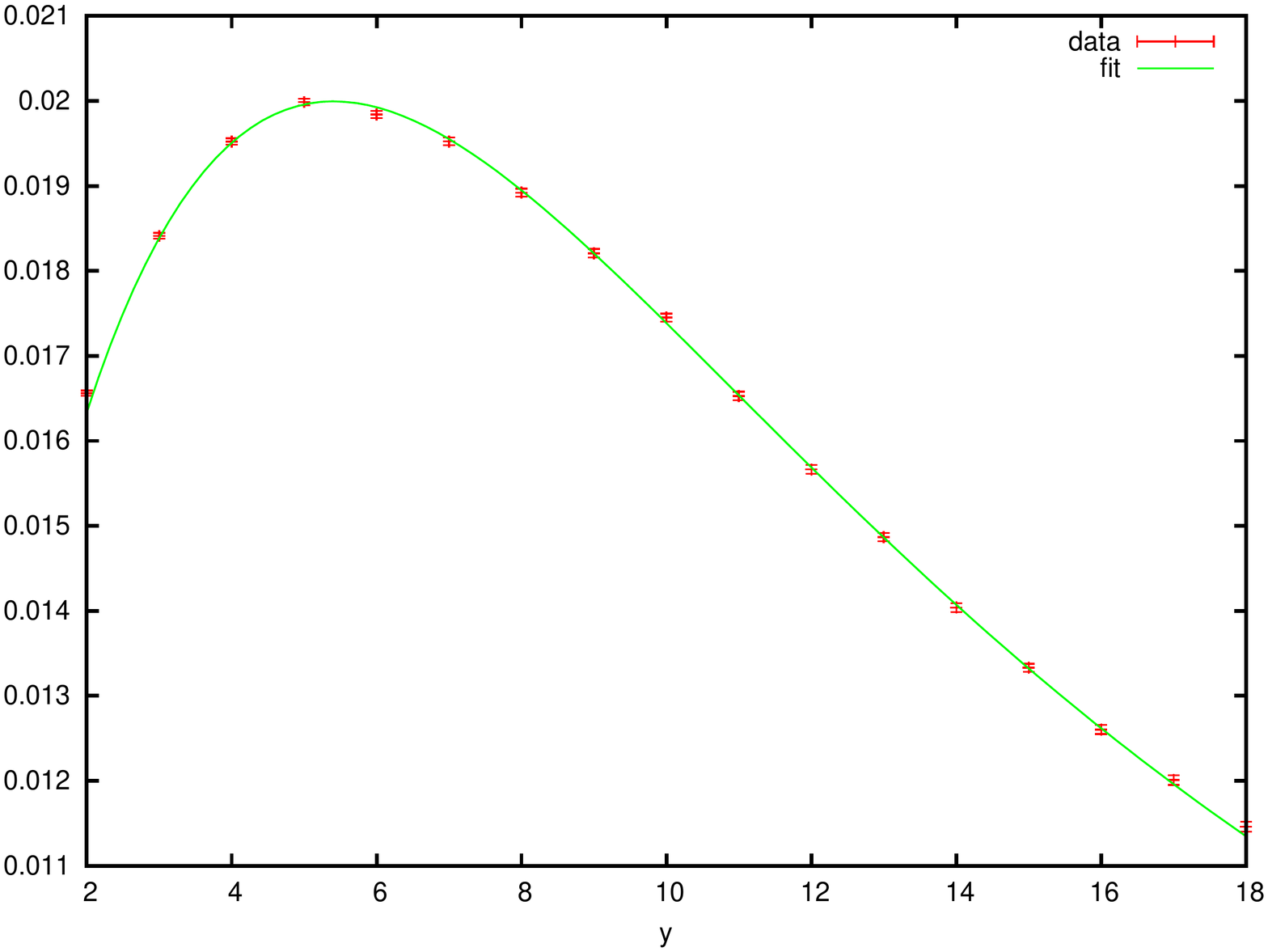}
\caption{The correlation function between the scalar spin on the bulk and a 
$J=\frac12$  defect local operator in a cubic lattice of size $\ell^3$ with 
$\ell=120$ with periodic%
 boundary conditions. The solid line is a one-parameter fit to (\ref{mixedJ}) 
with $d=10$. 
In this fit we also considered  the contribution of the nearest copy of 
the defect line, 
see the discussion in footnote \ref{foot2}.  
}
\label{fig:mixed_corr}
\end{figure}

We also considered mixed correlation functions of local operators in the bulk and on the defect as discussed in Eq. (\ref{mixed}). In particular, we placed at the origin a 
defect operator $\mathfrak{o}_J$ of spin $J$ 
and took as bulk operator the spin $\sigma$ associated to  a node at a 
distance $|x^i|=r$ from the defect line and at a distance $|x^\mu|=\sqrt{d^2+r^2}$ from the origin, see Fig. \ref{fig:mixed_setup}. In this case if the critical Ising model is a conformal-invariant theory we expect that
 \begin{equation}
 \langle \sigma(d,r,\phi) \mathfrak{o}_J(0) \rangle = C_J\, \ee^{-\ii\phi J} 
\frac{r^{\Delta_{\mathfrak{o}_J}-\Delta_{\sigma}}}{(r^2+d^2)^{\Delta_{\mathfrak{o}_J}}}\,,
\label{mixedJ}
\end{equation}
where $\Delta_\sigma=0.5182(2)$ 
is  the anomalous dimension of $\sigma$, while
$\phi$ is the azimuth angle around the defect line, with $\phi=0$ 
corresponding to the surface of frustrated links. The setup drawn in 
Fig. \ref{fig:mixed_setup} corresponds to $\phi=\pi$. 
In our numerical calculations 
we observed the best signal at $J=1/2$, with the operator $\mathfrak{o}_{1/2}$ corresponding to $\psi$. Notice that Eq. (\ref{defh1}) fixes completely the  phase of this mixed correlation function. 
In fact we have
\begin{equation}
\arg( \langle \sigma(d,r,\phi) \psi(0) \rangle)=
\arg(1+\omega^3)+\frac{\pi-\phi}2=\frac78\pi-\frac\phi2\,.
\end{equation}   
Similarly, if we considered the operator $\mathfrak{o}_{3/2}=\psi_{3/2}$ we would have 
\begin{equation}
\arg( \langle \sigma(d,r,\phi) \psi_{3/2}(0) \rangle)=
\arg(1+\omega)+\frac{3\pi-3\phi}2=\frac{13}8\pi-\frac{3\phi}2\,.
\end{equation} 
We plot in Fig. \ref{fig:mixed_corr} the numerical data of the 
imaginary part of  $\langle \sigma(d,r,\pi) \psi(0) \rangle$ taken at 
fixed longitudinal distance $d=10$, as well as its one-parameter fit%
\footnote{\label{foot2}In comparing our numerical results on the torus to eq. (\ref{mixedJ}), written on the covering space, we have to take into account the fact that in the covering space there are replicas of the defect lines. The replica of the $J=1/2$ defect operator closest to the bulk operator $\sigma$ is the one
that would appear at distance $\ell$ in the vertical direction from the one drawn in Fig. \ref{fig:mixed_setup}. 
With respect to it, the bulk spin is at transverse distance $r'=\ell-r$ and at an angle $\phi'=0$;
we find that the corresponding contribution of the form (\ref{mixedJ}) has to be included in the fit, while further replicas give neglegible contributions.}
to (\ref{mixedJ}).  

Another observable we studied in our simulations is the one-point function of  the energy operator in presence of a line defect. This can be considered as 
a mixed bulk-defect correlation function in the case in which the defect 
operator is the identity. In this particular case $\Delta_\mathbf{1}=0$
and  Eq. (\ref{mixed}) gives
\begin{equation}
\vev{\epsilon(x)\,\mathbf{1}(0)}=\frac{C^\epsilon_{\mathbf{1}}}{\vert x^i\vert^{\Delta_{\epsilon}}}\,,
\label{energy}
\end{equation}
where  $\vert x^i\vert$ is  the distance from the defect, 
while $C^\epsilon_{\mathbf{1}}$ is a numerical coefficient which ``measures'' the bulk-to-defect pairing of the energy operator, which is normalized in the bulk in the standard way, i.e. 
$\langle\epsilon(x)\epsilon(0)\rangle= \vert x_\mu\vert^{-2\Delta_\epsilon}$, with
$\Delta_{\epsilon}=1.4130(5)$. 

In our 
simulations we used as a probe  the link variable 
$\sigma_i\sigma_j$, which can be decomposed as the sum of the identity 
and the tower of local energy operators $\epsilon,\epsilon',\dots$. If this probe is sufficiently far from the defect line only the operators of lowest dimension contributes; thus we expect to have, in presence of  a defect line,
\begin{equation}
\vev{\sigma_i\sigma_j(x)}_{\rm defect} = \vev{\sigma_i\sigma_j(x) \mathbf{1}(0)}
={\rm constant}~+
\frac{a_\epsilon}{\vert x^i\vert^{\Delta_\epsilon}}+{\rm higher~ order~terms}\,, 
\end{equation}
while in the bulk, in absence of the defect line, the correlator of two link variables $\sigma_i\sigma_j$ and $\sigma_k\sigma_l$ 
is
\begin{equation}
\vev{\sigma_i\sigma_j(x) \sigma_k\sigma_l(0)}={\rm constant}~+
\frac{c_{\epsilon\epsilon}}{\vert x^\mu\vert^{2\Delta_\epsilon}}+{\rm higher~ order~terms}~.
\end{equation}
The coefficient $C^\epsilon_{\mathbf{1}}$  of Eq. (\ref{energy}) is thus given by
the universal amplitude ratio
\begin{equation}
C^\epsilon_{\mathbf{1}}=\frac{a_\epsilon}{\sqrt{c_{\epsilon\epsilon}}}\,.
\end{equation}
\begin{table}[ht]
\centering
\begin{tabular}{|ccc|}
\hline
$C^\epsilon_{\mathbf{1}}$&$\vert C^\sigma_\psi\vert$&$\vert C^\sigma_{\psi_{3/2}}\vert$\\
\hline
-0.167(4)&0.968(2)&0.61(9)\\
\hline
\end{tabular}
\caption{Universal amplitude ratios defining the bulk-to-defect pairing associated with the energy operator $\epsilon$ and the scalar $\sigma$, defined in 
Eq. (\ref{energy}) and in Eq. (\ref{sigma}).}
\label{tab:universalratios}
\end{table}

The bulk-to-defect pairing of the scalar field $\sigma$ is encoded in the constant $C_J$ of Eq. (\ref{mixedJ}) which allows to define further universal ratios
\begin{equation}
\vert C^\sigma_{ \mathfrak{o}_J}\vert=\left\vert\frac{ C_J}{\sqrt{c_{\sigma\sigma}
a_{ \mathfrak{o}_J \mathfrak{o}_J}}  }\right\vert\,,
\label{sigma}
\end{equation}
where $c_{\sigma\sigma}$ and $a_{ \mathfrak{o}_J \mathfrak{o}_J}$ are the coefficients of the 2-point functions of the scalar $\sigma$ on the bulk and of the operator 
 $\mathfrak{o}_J$ on the defect line. In Table \ref{tab:universalratios} we report our estimates for these universal ratios for $J=1/2$ and $J=3/2$ as well as  that associated with the energy operator.

\section{Conclusions}
\label{sec:concl}
The first lesson we draw from our numerical results is that the hypothesis of conformal invariance of the monodromy defect in the critical 3d Ising model 
seems well supported. From a theoretical point of view, it would be useful to fully spell out the conditions for a generic scale invariant defect in a CFT to be conformal, generalizing the known results for boundary conditions
\cite{Cardy:1984bb}. It would also be interesting to test this assumption in other concrete examples. 

The second lesson is that numerical methods are suitable to derive information about the lowest lying operators on the defect, including both 
their anomalous dimensions and various OPE coefficients. Our results suggest it may be interesting to revisit the problem of computing numerically 
OPE coefficients and correlation functions in the bulk theory, and compare them with the results of bootstrap. 

The most immediate direction for future inquiry is to test our results against analytic and semi-analytic methods such as the conformal bootstrap and the $\epsilon$ expansion. 
Notice that the monodromy defect can be defined uniformly in a scalar theory with $\phi^4$ interaction in the whole interval of dimensions 
$2 \leq D < 4$ where the bulk theory is expected to have an infrared conformal fixed point.  It should thus be possible to compute both anomalous dimensions and 
correlation functions in $4-\epsilon$ and $2+\epsilon$ dimensions. 

We find interesting that the spin $1/2$ operator on the monodromy defect has dimension only slightly smaller than $1$, and is thus 
very weakly relevant when integrated on the defect. This raises the possibility of a fully perturbative RG flow to some nearby conformal fixed point without rotational symmetry. 
It would be interesting to investigate this flow. 

\section*{Acknowledgements}
\dgb The research of DG  was supported by the Perimeter Institute for Theoretical Physics. Research at Perimeter Institute is supported by the Government of Canada through Industry Canada and by the Province of Ontario through the Ministry of Economic Development and Innovation. \dge
The work of MB was supported in part by the MIUR-PRIN contract 2009-KHZKRX. MM thanks Ettore Vicari and Michele Mintchev for useful discussions.

\end{document}